\documentclass[12pt]{article}

\usepackage{amssymb,textcomp,amsmath,graphicx,color,bigints,multirow}
\DeclareMathAlphabet{\mathcal}{OMS}{cmsy}{m}{n}
\long\def\symbolfootnote[#1]#2{\begingroup
\def\thefootnote{\fnsymbol{footnote}}
\footnote[#1]{#2}\endgroup}

\numberwithin{equation}{section}

\newcommand{\ie}{{\em i.e.~}}

\newcommand{\etal}{\emph{et\,al.~}}

\newcommand{\be}{\begin{equation}}  \newcommand{\ee}{\end{equation}}
\newcommand{\bea}{\begin{eqnarray}} \newcommand{\eea}{\end{eqnarray}}
\newcommand{\nn}{\nonumber}
\DeclareMathOperator{\erf}{erf}
\DeclareMathOperator{\erfc}{erfc}

\newcommand{\bse}{\begin{subequations}}
\newcommand{\ese}{\end{subequations}}
\newcommand{\bi}{\begin{itemize}}
\newcommand{\ei}{\end{itemize}}
\newcommand{\lp}{\left(}
\newcommand{\rp}{\right)}

\newcommand{\bh}{\rm PBH}

\begin{document}

\begin{center}
\Large{Primordial Black Holes in the Excursion Set Theory}
\end{center}
\begin{center}
\large{Encieh Erfani$^{1,\,*}$}, \large{Hamed Kameli$^{2}$} and \large{Shant Baghram$^{2,\,\dag}$}
\end{center}
\begin{center}
\textit{$^1$Department of Physics, Institute for Advanced Studies in Basic Sciences (IASBS), Zanjan 45137-66731, Iran\\
$^2$Department of Physics, Sharif University of Technology, Tehran 11155-9161, Iran}
\end{center}

\date{}

\symbolfootnote[0]{$^{*}$erfani@iasbs.ac.ir}
\symbolfootnote[0]{$^{\dag}$baghram@sharif.edu}

\begin{abstract}
We study primordial black holes (PBHs) formation in the excursion set theory (EST) in a vast range of PBHs masses with and without confirmed constraints on their abundance. In this work, we introduce a new concept of the first touch in the context of EST for PBHs formation. This new framework takes into account the earlier horizon reentry of smaller masses. Our study shows that in the EST, it is possible to produce PBHs in different mass-range, with enhanced power spectrum, which could make up all dark matter. We also show that in a broad blue-tilted power spectrum, the production of PBHs is dominated by smaller masses. Our analysis put an upper limit $\sim\,$0.1 on the amplitude of the curvature power spectrum at length scales relevant for PBHs formation.
\end{abstract}

\newpage

\section{Introduction}
\paragraph{}

Primordial black holes (PBHs) can form from the collapse of large density fluctuations in the early Universe \cite{Zeldovich:1967lct, 1971MNRAS.152...75H, 1974MNRAS.168..399C}. When the density fluctuations larger than the threshold reenters the horizon after inflation, the region will collapse to form a PBH with a mass roughly equal to the mass of the horizon. According to the Hawking radiation \cite{Hawking:1974sw}, PBHs with mass larger than $\sim10^{15}$ g have a lifetime longer than the age of the Universe. Moreover, since they have formed before matter-radiation equality, they are non-baryonic. Thus they could be a candidate for dark matter (DM).

After the discovery of gravitational waves (GWs) in 2016 by LIGO/Virgo from mergers of tens of Solar mass black holes \cite{Abbott:2016blz}, the possibility that these BHs could be primordial rather than astrophysical \cite{Bird:2016dcv, Clesse:2016vqa}, has led to a great interest in PBH version of DM. In addition, the abundance of PBHs in our observable Universe can give us some clues of small-scale density fluctuations which are not accessible via observations of the cosmic microwave background (CMB) and the large scale structure (LSS) (\ie constraints on the primordial power spectrum for scales between $k \sim (10^{-3} - 1)$ Mpc$^{-1}$) \cite{Cole:2017gle, Sato-Polito:2019hws, Kalaja:2019uju}. (For recent reviews on PBHs, see also refs.~\cite{Sasaki:2018dmp, Carr:2020gox}.) The power spectrum at small scales relevant for PBHs formation must become orders of magnitude larger than $\mathcal{O}(10^{-9})$ detected on large scales via the CMB \cite{Akrami:2018odb} which is a matter of active research. Since the amplitude of the power spectrum is only logarithmically sensitive to the PBHs abundance during the radiation domination, this variation has little to do with the different PBH masses \cite{Green:2020jor, Young:2014ana}. However, the PBH mass distribution differs when using different theoretical techniques for their formation \cite{Gow:2020bzo}\footnote{For instance, primordial non-Gaussianity can have an important effect on the required power spectrum amplitude \cite{Franciolini:2018vbk}. However, we will not consider this issue in this paper.}. PBH's formation has been studied with the use of Press-Schechter (PS) formalism \cite{1974ApJ...187..425P}, peaks theory \cite{1986ApJ...304...15B, Germani:2019zez} and recently by excursion set theory (EST) \cite{MoradinezhadDizgah:2019wjf, Auclair:2020csm}. In this paper, we make the first detailed numerical study of the mass distribution of PBH in the EST \cite{1991ApJ...379..440B} by counting the number of trajectories that reenter the horizon corresponding to a specific mass.
We study the formation of PBHs when a broad spectrum characterizes the scalar perturbations, and the goal of the paper is to answer the question of the mass distribution of PBHs for broad spectra in the EST. We will show that despite the broadness of the blue-tilted power spectrum, the production of PBHs enhances towards small masses.

There are various constraints on PBHs abundance which have recently been compiled in \cite{Carr:2020gox, Green:2020jor}. In our study, we will consider these observational constraints in mass ranges if they exist. Otherwise, we will assume that PBHs can comprise all of the DM. We will find the required blue-tilled spectral indexes for these mass ranges in the EST formalism. Finally, we will translate these results into upper limit on the amplitude of the curvature power spectrum at length scales relevant for PBHs formation.

The paper is laid out as follows: in section~\ref{sec2} we elaborate PBH formation in PS formalism, and we review current constraints on their density being associated with a variety of gravitational lensing \cite{Niikura:2017zjd, Niikura:2019kqi, Tisserand:2006zx, Allsman:2000kg} and GW \cite{Authors:2019qbw} effects. In section~\ref{sec3} we briefly explain the EST method and we develop the PBHs formation in this context. Finally, section~\ref{sec4} is devoted to our conclusions.

\section{Primordial Black Holes}\label{sec2}
\paragraph{}

The most common mechanism for PBHs formation is the collapse of large density perturbations generated by inflation in the very early Universe. These density perturbations that reenter the Hubble horizon in the radiation dominated (RD) era would gravitationally collapse into a BH if their amplitude is larger than a critical value. Since PBHs are not formed by stellar core collapse, they may be of any size from the Planck mass $\sim10^{-5}\,$g to $\sim10^{50}\,$g \cite{Carr:2020gox}. Mass of PBH, $M_{\bh}$ produced at a given time is limited by the horizon mass at that time
\be\label{mass}
M_{\bh} = \gamma\,M_{\rm PH} \sim 10^{15} \lp\dfrac{t}{10^{-23}\,{\rm s}}\rp {\rm g}\,,
\ee
where $M_{\rm PH}$ is the particle horizon mass and $\gamma$ is the fraction of the total energy that ends up inside the PBH\footnote{For simplicity we assume that the PBH mass is a fixed fraction of the horizon mass corresponding to the smoothing scale; $\gamma\simeq w^{3/2}\simeq 0.2$ during the RD era ($w=1/3$) \cite{Carr:1975qj}.}. Since PBHs radiate thermally \cite{Hawking:1974sw}, they evaporate on a time scale
\be\label{lifetime}
\tau_{_{\bh}}(M) \sim10^{64}\,\lp\frac{M}{M_{\odot}}\rp^{3}\,\rm{yr}\,,
\ee
therefore, the ones with mass greater than $10^{15}$ g survived until now and would be plausible DM candidates. In this paper, we will focus on DM PBHs formed in the RD era.

In order to investigate the abundance of formed PBHs, we define a parameter which represents the mass fraction (the energy density fraction) of PBHs,
\be\label{beta1}
\beta \equiv \frac{\rho_{\bh}(t_{\rm i})}{\rho(t_{\rm i})}\,,
\ee
where the subscript ``i'' indicates values at the epoch of PBH formation.
It is straightforward to show that this fraction can be related to present day abundance of PBHs, $f_{\bh}$ \cite{Carr:2009jm}
\be\label{beta2}
\beta \simeq 3.7\times10^{-9}\,\gamma^{-1/2}\,\lp\frac{g_{*,\,i}}{10.75}\rp^{1/4}\lp\frac{M_{\bh}}{M_{\odot}}\rp^{1/2} \,f_{\bh}\,,
\ee
where $g_{*,\,i}$ is a number of relativistic degree of freedom at the time of formation. Note that $f_{\bh}\equiv\Omega_{\bh}/\Omega_{\rm DM}$ is a fraction of PBHs against the total DM component, $\Omega_{\rm DM}$. Thus, for each mass of PBHs, the observational constraint on $f_{\bh}$ can be interpreted as that on $\beta$.\\
The fraction of the energy density of the Universe contained in regions overdense enough to form PBHs is usually calculated in PS theory \cite{1974ApJ...187..425P} as
\be\label{PS1}
\beta = \gamma \int_{\delta_c}^{\infty} P(\delta;\,R)\,d\delta\,.
\ee
Here $P(\delta;\,R)$ is the probability distribution function of the linear density field $\delta$ smoothed on a scale $R$, and $\delta_c$ is the critical threshold for PBHs formation. The estimation of $\delta_{c}$ is under discussion by different researches both analytically \cite{Harada:2013epa} and numerically \cite{Musco:2012au, Harada:2015yda}. In general, the value of $\delta_{c}$ is not unique and depends on the shape of the primordial curvature power spectrum \cite{Musco:2018rwt}.
The value of $\delta_{c}$ is varying between 0.4 and 2/3. In this paper, we consider $\delta_{c}=0.47$ which is the upper bound of the threshold calculated in \cite{Musco:2018rwt} for a Gaussian curvature profile \footnote{Note that in general, the value of $\delta_c$ is a function of the equation of state parameter \cite{Musco:2012au}. For example, a change in the relativistic degrees of freedom at the electroweak and QCD phase transitions induces a change in $\delta_c$. This will have unavoidable features in the mass function \cite{Byrnes:2018clq, Carr:2019kxo}.}.
For the Gaussian fluctuations, $\beta$ is given by \cite{Drees:2011hb}
\begin{align}\label{PS2}
\beta(M_{\bh}) & = \gamma \int_{\delta_c}^{\infty} \frac{d\delta}{\sqrt{2\pi\,\sigma^2_\delta(R)}} \, \exp\lp-\frac{\delta^2}{2\,\sigma^{2}_\delta(R)}\rp \nn\\
&= \frac{\gamma}{2}\,\erfc\lp\frac{\delta_c}{\sqrt{2\sigma^{2}}}\rp\,,
\end{align}
where $\erfc(x) = 1 - \erf(x)$ is the complementary error function and, the variance of $\delta$ is given by
\be\label{sigma}
\sigma^2_\delta (R) = \int \dfrac{dk}{k}\,\mathcal{P}_{\delta}(k)\,\widetilde{W}^2(k,\,R) = \int \dfrac{dk}{k}\,\dfrac{16}{81}\lp\dfrac{k}{aH}\rp^{4}\,\mathcal{P_R}(k)\,\widetilde{W}^2(k,\,R)\,,
\ee
where $\widetilde{W}(k,\,R)$ is the Fourier transform of the window function used to smooth the density contrast on a comoving scale $R$. In this paper, we will consider the $k$-space top-hat window function. Recall that $\mathcal{P}_{\delta}(k)$ is the power spectrum of $\delta$ which is related to the power spectrum of curvature perturbations on comoving hypersurfaces in RD era as follows \cite{Lyth:2009zz}
\be\label{2p}
\mathcal{P}_{\delta}(k) = \dfrac{16}{81}\lp\dfrac{k}{aH}\rp^4\mathcal{P_R}(k)\,,
\ee
and we parameterize the curvature power spectrum as
\be\label{power}
\mathcal{P_R}(k) = A_0\,\lp\frac{k}{k_{0}}\rp^{n_s(k) - 1}\,,
\ee
where $\ln(10^{10}A_0) = 3.044 \pm 0.014$ and the spectral index, $n_s(k_0)= 0.9649 \pm 0.0042$ are known by the CMB observation at $k_0=0.05$ Mpc$^{-1}$ \cite{Akrami:2018odb}.\\
As already mentioned, the mass of PBH, $M_{\bh}$ is related to scale of horizon at the time of formation when the perturbation reenters the horizon ($k = aH$). It is straightforward to obtain the relation between the mass of PBHs and the comoving wavenumbers as \cite{Kawasaki:2016pql},
\be\label{M-k}
M_{\bh}(k) \simeq 30\,M_{\odot}\,\lp\dfrac{\gamma}{0.2}\rp\,\lp\dfrac{g_{*,\,i}}{10.75}\rp^{-1/6}\lp\dfrac{k_{\bh}}{2.9\times 10^5\,{\rm Mpc}^{-1}}\rp^{-2}\,.
\ee

If the power spectrum of curvature perturbations were scale invariant, using Eq.~\eqref{PS2}, and assuming $\delta_c \simeq 0.47$ and $\sigma_\delta^2 \propto A_0$, then the initial mass fraction of PBHs formed in the RD era would be completely negligible. On the other hand, if we assume that all of the DM is in PBHs with $M_{\bh} \sim M_{\odot}$ (\ie $f_{\bh} = 1$) then, from Eq.~\eqref{beta2}, the initial PBH mass fraction must be $\beta \sim 10^{-9}$. This means that the required power spectrum on this scale (see Eq.~\eqref{M-k}) must be $\mathcal{P_R}({k_{\bh}}) \sim 10^{-2}$ which is 7 orders of magnitude larger than the measured value on cosmological scales $(A_0 \sim10^{-9})$. Thus for DM PBHs formation, the amplitude of power must be enhanced exponentially in specific wavenumbers corresponding to PBHs mass. One way to enhance the power is to have a blue-tilted spectral index in the scale of PBHs formation, $n_{s(b)}(k_{\bh})$. Hence the abundance of PBHs with different masses (formed in different scales) could put constraints on the power spectrum beyond the range probed by cosmological observations.

There are various constraints on PBHs abundance. In this paper, we are interested in PBHs with a mass range from $\sim(10^{16} - 10^{46})$ g. Part of this mass range is probed by the gravitational lensing and the GWs observations. However, there remain mass windows where PBHs may constitute the whole DM. We review the potential constraints as follows:
\bi
\item\textbf{Gravitational Lensing:}
Long-duration microlensing is a significant probe to find the contribution of compact objects as DM.
These constraints have been acquired by a) Subaru HSC (Hyper Suprime-Cam) survey observing the microlensing of stars in M31 (Andromeda galaxy) \cite{Niikura:2017zjd}, b) OGLE (Optical Gravitational Lensing Experiment) microlensing survey of the Galactic bulge \cite{Niikura:2019kqi}, and c) EROS/MACHO \cite{Tisserand:2006zx, Allsman:2000kg} which monitors millions of stars in Magellanic Cloud, respectively.\\
- $10^{-11}\,M_{\odot} \lesssim M_{\bh} \lesssim 10^{-6}\,M_{\odot}$ with $f_{\bh}\sim 10^{-3}\,,$\\
- $10^{-6}\,M_{\odot} \lesssim M_{\bh} \lesssim 10^{-3}\,M_{\odot}$ with $f_{\bh} \sim 10^{-2}\,,$\\
- $10^{-3}\,M_{\odot} \lesssim M_{\bh} \lesssim 10^{-1}\,M_{\odot}$ with $f_{\bh} < 0.04$\,.

Note that if $f_{\rm PBH}$ is not negligible and $M_{\rm PBH}\sim$ stellar mass, the Poissonian noise due to the discrete nature of PBHs implies a boosted formation of PBH at high redshift, compared to the standard model \cite{Kashlinsky:2016sdv}. In turn, when most PBHs are in massive halos, they can evade the microlensing limits due to the additional lensing effect of the cluster, as shown by \cite{Carr:2019kxo}. However, in this work, we evade these complications and use the constraints mentioned by \cite{Carr:2020gox}.

\item\textbf{Gravitational Waves:}
The merger of sub-Solar PBH binaries emits GWs like those observed by LIGO/Virgo \cite{Authors:2019qbw}. The observed merger rates of the second LIGO/Virgo run impose constraints on PBH fraction in the following mass range\\
- $0.2\,M_{\odot} \lesssim M_{\bh} \lesssim 1.0\,M_{\odot}$, corresponding to at most $0.16$ or $0.02$ of the DM, respectively \footnote{For the Poisson noise, the limits from GWs and BH merging rates still allow $f_{\bh} >  0.1$ \cite{Raidal:2018bbj}.}.\\ 
The main constraints on solar mass PBH are GW and lensing, which both are model-dependent. For example, one can evade the lensing constraints by clustering \cite{Carr:2019kxo}. Therefore, we consider the optimistic fraction $f_{\text{PBH}}=1$ as well.\\
- intermediate mass range, $10\,M_{\odot} \lesssim M_{\bh} \lesssim 10^3\,M_{\odot}$: In this mass range there are several limits such as ultra-faint dwarf galaxies, X-rays  and radio binaries, wide binaries, and CMB distortion limits \cite{Carr:2020gox}. Thus we choose the average limit of $f_{\bh}\sim 10^{-4}$.
\item
For the following mass ranges there is no (model independent) confirmed constraints. Therefore, PBHs could be the whole DM, \ie $f_{\bh}\sim 1$ \cite{Carr:2020gox}. \\
- asteroid mass range $10^{16}\,{\rm g} \lesssim M_{\bh} \lesssim 10^{17}$ g\,,\\
- sublunar mass range $10^{20}\,{\rm g} \lesssim M_{\bh} \lesssim 10^{24}$ g\,,\\
- stupendously large black holes (SLABs) $M_{\bh} \geq 10^{11}\,M_{\odot}$ \cite{Carr:2020erq}.
\ei
Thus, in this paper, we will consider eight mass ranges. It is worth to mention that we only consider the constraints without any assumption about the distribution of PBHs at their formation, their clustering and the environment since these conditions can have a significant effect in their abundances.

\section{Abundance of PBH in the Context of EST}\label{sec3}

As already mentioned in section~\ref{sec2}, the PS formalism can determine the abundance of PBHs. In this section, we will consider this issue in EST as a sophisticated development of PS formalism. In subsection~\ref{sec3.1}, we will review the main idea of EST, and in subsection~\ref{sec3.2}, we will have our main results on studying the abundance of PBHs in this context.

\subsection{Theoretical Background}\label{sec3.1}

\begin{figure*}[h!]
\includegraphics[scale=1]{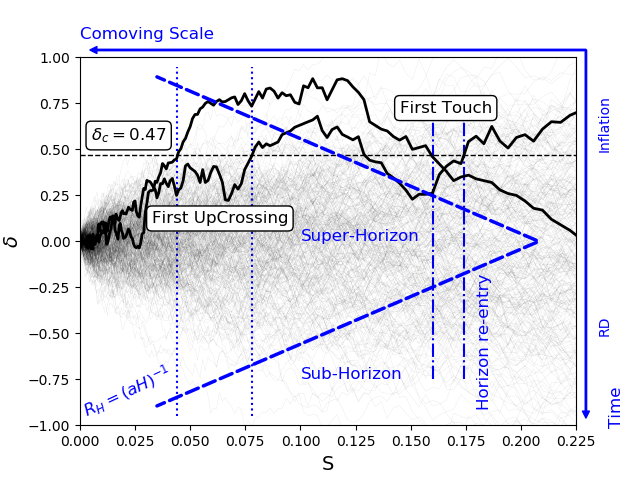}
\caption{The Markov trajectories were performed by the smoothed density contrast. The trajectories which pass the threshold, $\delta_c$ for the first time are called a) First Up-Crossing (FU): at lower $S$ (larger mass) b) First Touch (FT): at higher $S$ (smaller mass).
The blue inverted coordinates show the comoving scale versus cosmic time. The dashed inclined lines show the dependence of the comoving horizon $\lp R_H=(aH)^{-1}\rp$ to scale and time. The dash-dotted and dotted blue vertical lines correspond to the perturbation modes reentering the horizon in radiation dominated (RD) era, corresponding to FT and FU concepts, respectively.}
\label{trj1}
\end{figure*}

\begin{figure*}[h!]
\includegraphics[scale=1]{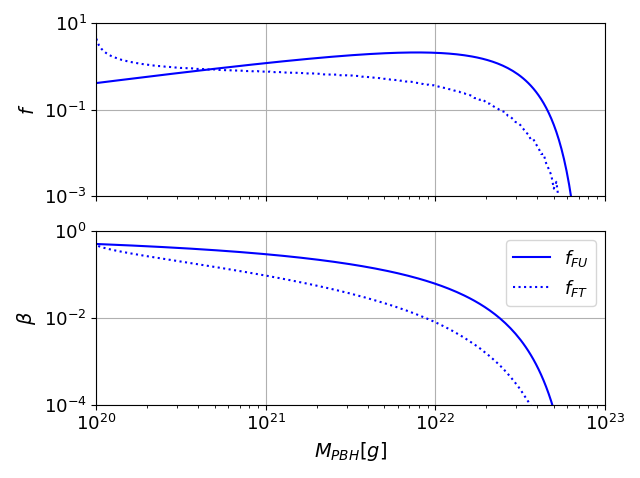}
\caption{Top panel: The Schematic  illustration of both FU (solid line) and FT (dotted line) for $f$ (fraction)  versus mass is plotted.  Bottom panel: The corresponding cumulative fraction, $\beta$ versus mass for both FU (solid line) and FT (dotted line) is shown.}
\label{fig-FTFU}
\end{figure*}

According to the EST, the spherical/ellipsoidal collapse of a compact object of mass $M$ forms where the corresponding smoothed linear density contrast, $\delta$ is larger than critical value, $\delta_{c}$ \cite{2010gfe..book.....M}. In this framework, the computation of the mass function of a compact object is formulated in terms of a stochastic process. In other words, if we smooth the density contrast in an initial box (higher redshifts where perturbations are almost linear and Gaussian) and plot the density contrast in terms of variance of perturbations $S$, it executes a stochastic time series.
Since $\delta$ is a function of variance, its evolution is governed by Gaussian white noise which implies the random walk with respect to $S$. The trajectories start from an initial value $\delta=0$ at $S=0$ (see Fig.~\ref{trj1}).
In this context, the first up-crossing (FU) trajectories that pass $\delta_c$ will give the mass of the collapsed object. Accordingly, the statistics of FU trajectories will give the distribution of collapsed objects. If we use the $k$-space top-hat window for smoothing, the random walks become Markovian and the distribution of FU trajectories is given by
\be\label{FU}
f_{\rm FU}(S,\,\delta_c) = \dfrac{1}{\sqrt{2 \pi}}\,\dfrac{\delta_c}{S^{3/2}} \exp\lp-\frac{\delta^{2}_{c}}{2S}\rp\,,
\ee
where the variance of density perturbations, $S$ is given by Eq.~\eqref{sigma}.\\
We will produce Markov trajectories by the method developed in our previous works \cite{Nikakhtar:2018qqg, Kameli:2019bki, Kameli:2020kao} where the distribution of DM halos in the (extended) standard model are addressed in more detail.\\
It is worth mentioning that there is a fundamental difference between the statistics of DM halos and PBHs. PBHs form when the scale corresponding to their mass reenters the horizon. In Fig.~\ref{trj1}, the vertical lines represent the modes that reenter the horizon on a specific scale. We should consider the trajectories which touch the barrier in larger variance for the first time, hereafter {\it first touch (FT)}. This approach is reasonable since the scale of PBH with lower mass (larger variance) reenters the horizon sooner than the massive one. The concept of FT is clear from Fig.~\ref{trj1}, where the FT trajectories touch the threshold in larger variance; \ie smaller mass PBHs form first. For more clarification, compare the vertical dashed-dotted lines for small mass PBH corresponding to FT with vertical dotted lines for larger masses corresponding to FU. Accordingly, the number of FT trajectories at each variance will give the abundance of PBHs with corresponding mass. Note that according to correspondence between PS and EST, the cumulative summation of FU distribution (Eq.~\eqref{FU}) is equal to the fraction of PBHs at their formation, $\beta$ (see Eq.~\eqref{PS1}). We show that by knowing the FU distribution for any given power spectrum, one can find $\beta$ and $f_{\bh}$ (see Eq.~\eqref{beta2}). The fraction of PBHs is given by counting the FT trajectories. However, statistically, the FT distribution is a correction to the FU ones with almost similar behaviour. To clarify this issue, in the schematic Fig.~\ref{fig-FTFU}, we plot the FU and FT distribution (top plot) and their cumulative distribution $\beta$ (bottom plot) for an arbitrary mass range. FT distribution is larger than FU at smaller masses and vice versa.
One could anticipate this result as the FT is related to the crossings in the smaller mass ranges in comparison to FU.
The cumulative distribution, $\beta$ of both FT and  FU is almost the same for the lower bound of the mass range.

\subsection{PBH in the Context of EST}\label{sec3.2}

To calculate the abundance of PBHs at different scales, the distribution is confined with a window function to smooth out modes smaller than the horizon. In the EST formalism for halo formation, the variance could be calculated at redshift $z=0$ and density threshold, $\delta_c$ changes with growth function, $D(z)$. However in the case of PBH, we use a constant $\delta_c=0.47$ in RD era \cite{Musco:2018rwt} and the power spectrum is identified in various horizon crossing scales, $R_H = (aH)^{-1}$. This means the redshift dependence is encapsulated in power spectrum instead of $\delta_c$. So we have a unique power spectrum for each horizon scale $R_H$ which is related to a specific mass. Therefore, Eq.~\eqref{2p} at horizon crossing scale $R_H$ is given by
\be\label{pow}
\mathcal{P}_{\delta}(k,\,R_H) \equiv \mathcal{P}_{\delta,R_H}(k) = \dfrac{16}{81}A_0\,\lp k R_H\rp^4 \lp\dfrac{k}{k_0}\rp^{n_s(k)-1}\,.
\ee
We use a variable spectral index to apply a blue-tilted power spectrum in a broad interval $k\sim [k_s,\,k_l]$ where $k_s$ and $k_l$ are short and long  wavenumbers. This means
\be\label{n}   
 n_s(k) = 
     \begin{cases}
         n_s(k_0)\,,     &  k < k_s\\
         n_{s(b)}>1\,,  &  k_s < k < k_l \\
         n_s(k_0)\,,    &   k > k_l\\ 
     \end{cases}
\ee
\begin{figure*}[h!]
\includegraphics[scale=1]{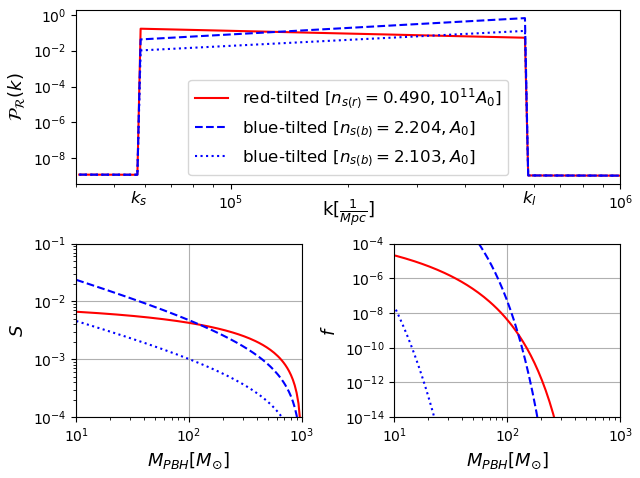}
\caption{Top panel: The enhanced initial power spectrum is plotted versus wavenumber. The enhancement is for blue-tilted (dashed and dotted lines) and red-tilted with $A=10^{11}A_0$ (solid line) models in wavenumber range of $k\sim [k_s,\,k_l]$. Bottom panel: The variance, $S$ and the fraction, $f$ versus PBH mass is plotted in left and right panels, respectively. }
\label{fig3}
\end{figure*}
where $n_s(k_0)= 0.9649 \pm 0.0042 $ and $n_{s(b)}$ are the observed spectral index and the blue-tilted one, respectively.
In Fig.~\ref{fig3}, top panel, we show the blue-tilted power enhancement in the wavenumber interval $k\sim [k_s,\,k_l]$.
Since we have a whole sequence of $S$ for different scales, $R_i$ at any specific horizon crossing scale $R_H$, we rewrite Eq.~\eqref{sigma} as follows
\be\label{var2}
S_{R_H}(R_i) = A\,R_H^4 \int_{0}^{\infty} dk\,k^{n_s(k)+2}\,\widetilde{W}^2(k,\,R_i)\,,
\ee
where $A \equiv \dfrac{16}{81}\dfrac{A_0}{k_0^{n_s(k)-1}}$. The above equation means that using each sequence of $S$, we form a whole realization of trajectories in $(\delta-S)$ plane for each $R_H$. Hence, we could count the FT numerically, \ie $f_{\rm FT,\,R_H}$. Therefore, for PBHs which form at horizon crossing, $f_{\rm FT,\,R_H}$ is only meaningful where the FT is counted at $R_i=R_H$.\\
Note that since PBHs formation at horizon reentry is independent of other scales, therefore their formation mechanism is history independent. In the EST, one can get memory independent trajectories by using the $k$-space top-hat window function which is known as Markov model\footnote{Note that the non-Markov extension of EST has great importance in halo formation due to more physical smoothing window functions which lead to correction of statistics of FUs, especially in small scales \cite{Nikakhtar:2018qqg, Kameli:2019bki, Baghram:2019jlu}.}.
For Markov model Eq.~\eqref{var2} is given by
\be\label{Sf}
\begin{aligned}
\frac{S_{R_H}(R_i)}{A\,R_H^4} & = \frac{1}{R_i^{n_{s}+3}}\int_{0}^{min(\frac{R_i}{R_s},\,1)}{dk\,k^{n_{s}+2}}\\
                                                  &\,+\,\frac{1}{R_i^{n_{s(b)}+3}}\int_{min(\frac{R_i}{R_s},\,1)}^{min(\frac{R_i}{R_l},\,1)}{dk\,k^{n_{s(b)}+2}}\\
                                                  &\,+\,\frac{1}{R_i^{n_{s}+3}}\int_{min(\frac{R_i}{R_l},\,1)}^{\infty}{dk\,k^{n_{s}+2}}\,.
\end{aligned}
\ee
The $k$-space window function applies sharp conditions in upper/lower limits of the integral.\\
Note that due to random nature of these trajectories, we only need to apply one sequence of $S$ and substitute $S(R_H) \equiv S_{R_H}(R_i=R_H)$. We can analytically calculate $f_{\rm FU}$ and count the first touch of trajectories to evaluate $f_{\rm FT}$. In counting process, we will exclude any trajectory which is already collapsed to a PBH.
\begin{figure*}[h!]
\includegraphics[scale=1]{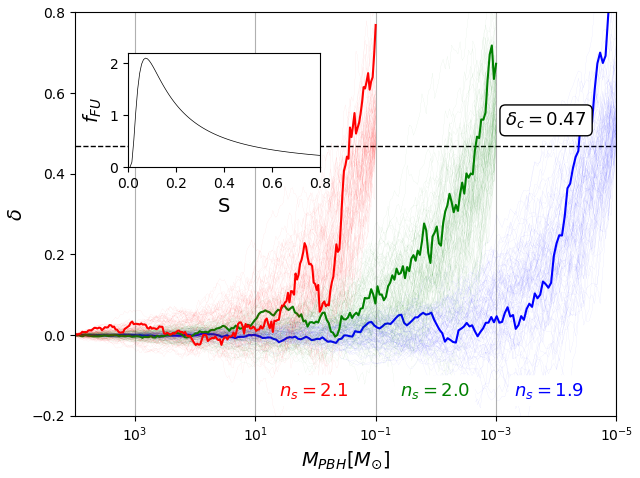}
\caption{Different trajectories are shown for different mass ranges. Higher masses require larger spectral index for their formation. In any mass range, the number of trajectories that pass the threshold is higher towards the smaller mass. The FU distribution (Eq.~\eqref{FU}) is shown in the inset figure.}
\label{trj2}
\end{figure*}
PBH formation needs amplification in primordial power spectrum. We modify the initial power spectrum introduced by early universe physics in a small wavenumber, where we do not have the constraints of CMB and LSS. We model this amplification by phenomenological enhancement. In Fig.~\ref{fig3}, we show the red and blue-tilted enhanced curvature power spectrum in wavenumber range of $k\sim [k_s,\,k_l]$ for a specific case. Note that for the red-tilted power enhancement, one needs an amplitude of an order $10^{11}A_0$. We also show that the variance and FU statistics change due to these two types of modifications are almost in the same order of magnitude (bottom panels for $S$ and $f$ versus mass) because the first crossing statistics depend only on the variance. The variance is a monotonic decreasing function of mass (scale), which is obtained by an integral over the power spectrum (see Eqs.~\eqref{n} and \eqref{var2}). In what follows, we use the blue-tilted power spectrum for PBHs formation, where the modification in spectral index requires the desired amplification in the mass range more naturally with the same amplitude of $A_0$. Therefore, we need to enhance the power spectrum on a specific scale (see Eq.~\eqref{Sf}). In Fig.~\ref{trj2}, we show three schematic realizations of trajectories corresponding to three different blue-tilted spectral indexes at different mass ranges (different variances).
Although all realizations of trajectories have the same $f_{\rm FU}$ distribution versus $S$ (see Eq.~\eqref{FU} and inset Fig.~\ref{trj2}), however, the difference between realizations is due to the variation of $S$ at different masses. These realizations also show that a larger spectral index leads to a higher PBH mass. In each mass range, the number of trajectories that pass the threshold is higher in the smaller mass range.
Note that although $f_{\rm FU}$ has a maximum in variance ($S_{max}$), however, the maximum fraction of PBHs are formed in declining part of the $f_{\rm FU}$ which justified by the $f_{\bh}$ constraints proposed in table~\ref{table}.

Hence, the obtained value of $S$ from Eq.~\eqref{Sf} will be larger than variance, which makes the fraction maximum. This calculated $S$ corresponds to a specific blue-tilted spectral index required for PBH formation (see Fig.~\ref{trj2}). We should note that $S$ is highly sensitive to spectral index, therefore a small increase (decrease) in $n_{s(b)}$ leads to a larger (smaller) mass PBH. For stupendously large masses \cite{Carr:2020erq} the obtained value of $S$ could be close to $S_{max}$.

We report the results of our studies for eight desired mass ranges in table~\ref{table}. In each mass range, we consider intervals with different lower limits because the statistic of FU/FT is higher for the lower mass of the interval. In this table, $f_{\bh}$ is based on observational constraint if there exist, otherwise, we suppose that PBHs comprise the whole of DM. We report the required value of blue-tilted spectral indexes for both EST and PS formalism for comparison. One can see that these values are close to each other as expected.

\begin{figure*}[h!]
\includegraphics[scale=.55]{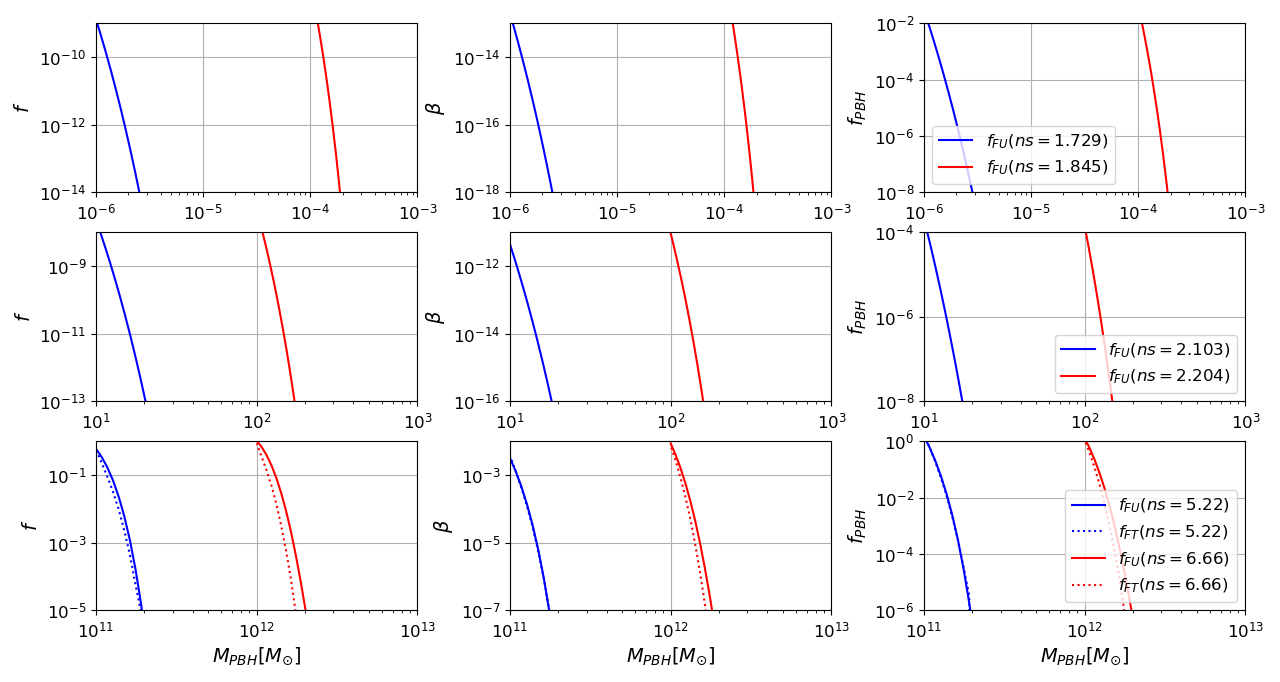}
\caption{From left for right: $f_{\rm FU/FT}$, $\beta$, and $f_{\bh}$ with respect to the mass of DM PBHs. From top to bottom: OGLE, intermediate and SLAB mass ranges. Note that $f_{\bh}$ is the required constraint for each mass range reported in table~\ref{table}.}
\label{final}
\end{figure*}

\begin{table*}[t!]
\begin{center}
\resizebox{\linewidth}{!}{
\begin{tabular}{ l | c | c | c | c | c |}

\cline{2-6}
& \multirow{2}{*}{mass range} & \multirow{2}{*}{$f_{\bh}$} & lower bound & \multicolumn{2}{|c|}{spectral index} \\ \cline{5-6}
& & & of mass range & EST & PS\\
\cline{1-6}
\multicolumn{1}{ | l | }{\multirow{1}{*}{asteroid mass range}} & $\lp10^{16} - 10^{17}\rp\,{\rm g}$ & 1 & $10^{16}$ g & 1.490 & 1.616 \\ \cline{4-6}
\cline{1-6}
\multicolumn{1}{ | l | }{\multirow{2}{*}{sublunar mass range}} & \multirow{2}{*}{$\lp10^{20} - 10^{24}\rp\,{\rm g}$} & \multirow{2}{*}{1} & $10^{20}$ g & 1.540 & 1.703 \\ \cline{4-6}
\multicolumn{1}{ |l | }{}
& & & $10^{22}$ g & 1.600 & 1.756\\
\cline{1-6}
\multicolumn{1}{ | l | }{\multirow{2}{*}{Subaru HSC}} & \multirow{2}{*}{$\lp10^{-11} - 10^{-6}\rp\,M_{\odot}$} & \multirow{2}{*}{$10^{-3}$} & $10^{-10}\,M_{\odot}$ & 1.604 & 1.560 \\ \cline{4-6}
\multicolumn{1}{ |l | }{}
& & & $10^{-8}\,M_{\odot}$ & 1.666 & 1.605\\
\cline{1-6}
\multicolumn{1}{ | l | }{\multirow{2}{*}{OGLE}} & \multirow{2}{*}{$\lp10^{-6} - 10^{-3}\rp\,M_{\odot}$} & \multirow{2}{*}{$10^{-2}$} & $10^{-6}\,M_{\odot}$ & 1.729 & 1.757\\ \cline{4-6}
\multicolumn{1}{ |l | }{}
& & & $10^{-4}\,M_{\odot}$ & 1.845 & 1.835 \\
\cline{1-6}
\multicolumn{1}{ | l | }{\multirow{2}{*}{EROS/MACHO}} & \multirow{2}{*}{$\lp10^{-3} - 10^{-1}\rp\,M_{\odot}$} & \multirow{2}{*}{$0.04$} & $10^{-3}\,M_{\odot}$ & 1.862 & 1.947 \\ \cline{4-6}
\multicolumn{1}{ |l | }{}
& & & $10^{-2}\,M_{\odot}$ & 1.942 & 1.970\\
\cline{1-6}
\multicolumn{1}{ | l | }{\multirow{4}{*}{sub-Solar mass range*}} & \multirow{4}{*}{$\lp0.2 - 1.0\rp\,M_{\odot}$} & \multirow{2}{*}{$0.02$} & $0.2\,M_{\odot}$ & 2.018 & 2.046 \\ \cline{4-6}
\multicolumn{1}{ |l | }{}
& & & $0.6\,M_{\odot}$ & 2.115 & 2.078 \\
\cline{3-6}
\multicolumn{1}{ | l | }{\multirow{2}{*}{}} &  & \multirow{2}{*}{$1$} & $0.2\,M_{\odot}$ & 2.028 & 2.258\\ \cline{4-6}
\multicolumn{1}{ |l | }{}
& & & $0.6\,M_{\odot}$ & 2.126 & 2.297 \\
\cline{1-6}
\multicolumn{1}{ | l | }{\multirow{2}{*}{intermediate mass range}} & \multirow{2}{*}{$\lp10^1 - 10^3\rp\,M_{\odot}$} & \multirow{2}{*}{$10^{-4}$} & $10\,M_{\odot}$ & 2.103 & 1.848\\ \cline{4-6}
\multicolumn{1}{ |l | }{}
& & & $10^{2}\,M_{\odot}$ & 2.204 & 1.911\\
\cline{1-6}
\multicolumn{1}{ | l | }{\multirow{2}{*}{SLABs}} & \multirow{2}{*}{$\geq 10^{11}\,M_{\odot}$} & \multirow{2}{*}{1} & $10^{11}\,M_{\odot}$ & 5.220 & 5.598\\ \cline{4-6}
\multicolumn{1}{ |l | }{}
& & & $10^{12}\,M_{\odot}$ & 6.660 & 6.891\\
\cline{1-6}
\end{tabular}}
\end{center}
\caption{Reported spectral indexes in EST and PS for eight mass ranges based on $f_{\bh}$ constraints. *\,Note that the sub-Solar mass range is studied with conservative limit of $f_{\bh}=0.02$ and optimistic limit $f_{\bh}=1$.}
\label{table}
\end{table*}

In Fig.~\ref{final}, we show the abundance of PBHs by counting both the FT and FU for OGLE, intermediate, and SLAB mass ranges from top to bottom, respectively. In each mass range, we show $f_{\rm FU/FT}$, $\beta$, and $f_{\bh}$ from left to right, respectively. Our numerical results show that despite considering a broad spectrum characterized by a range of scales running from $1/k_l$ to $1/k_s$, production of PBHs will be dominated by smaller masses (\ie smaller scales, $1/k_l$). For SLAB mass range we show both FU and FT, however for smaller mass ranges due to computational limits, we  only report FU. According to Eq.~\eqref{beta2}, $f_{\bh}$ is much larger than $\beta$ for small mass range, which is clear from the middle and the right plots of Fig.~\ref{final}. This means that to get the desired $f_{\bh}$ for smaller mass, we need very low FU/FT. And since we calculate FU analytically and FT numerically, small FT abundance requires enormous computational effort to produce enough FT trajectories. Since we have already shown in Fig.~\ref{fig-FTFU} that there is no considerable difference between FT and FU, therefore it is convenient to report only the results of FU for small mass ranges.

The constraints on the spectral indexes (see table~\ref{table}) can be translated to the upper limit on the power spectrum of curvature perturbations, $\mathcal{P_R}$ by using Eq.~(\ref{M-k}) to find wavenumbers corresponding to PBH masses. We obtain $\mathcal{P_R}(k_{\bh}) \lesssim 10^{-1}$ with some scale dependence in the range of $k_{\bh}\sim (10^{ 5} - 10^{15})\,\text{Mpc}^{-1}$. The power spectrum corresponding to the scale of SLABs ($k \sim (1-5)\,\text{Mpc}^{-1}$) is tightly constrained by Lyman-$\alpha$ \cite{Viel:2009ak}. Hence, the whole DM can not be made of SLAB PBHs.

\section{Conclusions}\label{sec4}

Although the Primordial black holes were first theorized decades ago, the first gravitational wave detection by LIGO/Virgo, the great interest towards PBHs as a DM candidate has revived. Consequently, there have been significant improvements in the theoretical calculations of PBH formation and the observational constraints on their abundance. Traditionally, the Press-Schechter (PS) formalism has been used for PBHs formation. In this work, we use the Excursion Set Theory (EST) as a reasonable and sophisticated extension of PS.
In the EST formalism, the trajectories which first up-cross (FU) the density threshold, $\delta_c$ (which changes with growth function), form a dark matter halo. However, for PBHs, we implement the EST with a constant density threshold and encapsulate the redshift dependence in the power spectrum. We also introduce a new concept of first touch (FT) instead of FU for the first time (see Fig.~\ref{trj1} and related discussion in section~\ref{sec3.1}).
Since PBHs form when the scale corresponding to their mass reenters the horizon, this leads to the use of the Markov trajectories as a memory-independent method. We showed in any mass range, the small PBHs form first and dominate the mass profile of PBHs. For this reason, we count FT at smaller masses instead of FU (see Fig.~\ref{fig-FTFU}).\\
Since PBHs formation is only possible by enhancing the power spectrum at small scales, we use a blue-tilted power spectrum in our analysis. We also show that a red-tilted power with a larger value of amplitude leads to almost similar results in PBH formation. Our numerical results showed that in a broad spectrum, the production of PBHs is dominated by smaller masses. We report the results of required blue-tilted spectral indexes in table~\ref{table} for eight mass ranges in the EST formalism. As a by-product, we compared EST with PS, and we showed that they are in good agreement in the considered masses.
The constraints on the spectral indexes were translated to the upper bound on curvature power spectrum, $\mathcal{P_R}(k_{\bh}) \lesssim 10^{-1}$ with some scale dependence in the range of $k_{\bh}\sim (10^{ 5} - 10^{15})\,\text{Mpc}^{-1}$.\\
It is worth mentioning that our approach could be applied for any redshift/scale dependence of the density threshold. For future work, we will apply the EST to study the merger history of PBHs and their redshift evolution. Also, it worths extending our work to non-Markov and non-Gaussian perturbations.

\section*{Acknowledgments}
We are grateful to Sohrab Rahvar for his insightful comments on the manuscript.
EE and SB are partially supported by Abdus Salam International Center of Theoretical Physics (ICTP) under the junior associateship scheme. They thanks the hospitality of ICTP where this work is initiated. We thank Nasim Derakhshanian for her collaboration in the early stage of this work. SB is supported by Sharif University of Technology Office of Vice President for Research under Grant No. G960202.

\end{document}